# Development of Habits Through Apprenticeship in a Community: Conceptual Model of Physics Teacher Preparation

Eugenia Etkina[1], Bor Gregorcic[2], and Stamatis Vokos[3]

[1] Graduate School of Education, Rutgers University, New Brunswick, New Jersey 08904, USA
[2] Department of Physics and Astronomy, Uppsala University, Box 516, 751 20 Uppsala, Sweden
[3] Department of Physics, Seattle Pacific University, Seattle, Washington 98119, USA

**Abstract:**
We propose the DHAC conceptual model of teacher preparation (Development of Habits through Apprenticeship in a Community), which builds on the contemporary literature on teacher preparation in physics and other disciplines and strives to provide a better understanding of the process of teacher formation. Extant literature on teacher preparation suggests that pre-service teachers learn best when they are immersed in a community that allows them to develop dispositions, knowledge, and practical skills and share with the community a strong vision of what good teaching entails. However, despite having developed the requisite dispositions, knowledge, and skills in pursuing the shared vision of good teaching, the professional demands on a teacher's time are so great out of and so complex during class time that if every decision requires multiple considerations and deliberations with oneself, the productive decisions might not materialize. We therefore argue that the missing link between intentional decision-making and actual teaching practice are the teacher's habits (spontaneous responses to situational cues). Teachers unavoidably develop habits with practical experience and under the influence of knowledge and belief structures that in many ways condition the responses of teachers in their practical work. To steer new teachers away from developing unproductive habits directed towards "survival" instead of student learning, we argue that any teacher preparation program (e.g., physics) should strive to develop in pre-service teachers strong habits of mind and practice that will serve as an underlying support structure for beginning teachers. These habits form the core of the conceptual model described in the paper. We provide examples of habits, which physics teachers need to develop before starting teaching and propose mechanisms for the development of such habits.

## I. INTRODUCTION

The purpose of this paper is to propose a conceptual model to guide the design of the professional preparation of teachers of physics. Our model elements describe both essential characteristics of good teachers and the requisite features of teacher preparation programs that have the development of those characteristics as their goal.

In this section, we argue for the *need* of a *general* theoretical framework to help organize conceptually the plethora of articles, reports, and books on effective STEM teacher education and professional development. Furthermore, we demonstrate how our proposed model, although probably applicable to almost any subject matter, is *articulated* and *instantiated* in a discipline-specific manner. Here we choose physics as our context because it is the only context that we know in detail and for which we may make claims. To say that a teacher education framework should reflect the characteristics, values, and practices of the discipline is important (and well understood in the literature), yet it is incomplete, as it does not highlight *which* characteristics, values, and practices are important for a specific discipline or how to foster their development. Because of the (very) small numbers of physics teachers prepared per U.S. institution [1], the current teacher preparation efforts almost always treat all science teachers (or even science and math teachers) interchangeably. For this reason, the formulation of our model privileges the *special* areas that need to be attended by universities to better serve the needs of *physics* teachers, and ultimately of *physics* itself.

*There is a need for theory*

There is extensive literature on characteristics of effective teachers. For instance, one may look at teacher formation and development through the lens of subject matter knowledge, or reflective practice, or teacher leadership traits, or the ability to engage in proximal formative assessment, or fluency in cultural responsiveness, or through value-added analyses of student performance on state high stakes tests, *etc*. Or one may appeal to accreditation standards [2], such as those of the National Board for Professional Teaching Standards [3], the Council for the Accreditation of Educator Preparation [4], or the American Association of Physics Teachers [5,6]. Taken together, a set of complex *patterns, or characteristics, or features of effective teacher preparation programs* emerge that are informed by extant practice in the field, classroom experience of program designers and implementers, and policy preferences.[1] Epistemologically, however, noticing a pattern is different than having a *model* that *explains* the pattern. To the best of our knowledge, there is no such conceptual model currently in the literature, at least as we use the term.

*There is a need for a discipline-specific model for teacher preparation.*

Having established the need for a conceptual model, one cannot help but ask whether one expects such a model to be sufficiently different in physics than, say, biology or secondary science more generally. Two major reasons compel us to put forth a physics-specific model for teacher preparation. First, we take as axiomatic that the research on the learning and teaching of the discipline should serve to guide the design and implementation of teacher education programs. The physics education research and non-physics discipline-based education research (DBER) enterprises have built overlapping but culturally different communities. Physics teacher preparation then should celebrate the distinctiveness of PER by capitalizing on the extant literature on physics conceptual understanding, problem-solving, laboratory work, instructional strategies, student expectations, student affect, *etc*. [7,8] Second, each disciplinary community has its own values, epistemological commitments, and particular ways of thinking. If we want teachers to serve first and foremost as ambassadors for physics, it follows that they should have a deep sense of cultural participation in the *specific* practices of physics.

---

[1] Although there may be significant overlap in the *patterns* that different policy designers privilege, there are profound disagreements in some circles about good ways to instantiate these patterns in teacher education programs. Some questions that are being debated in policy circles and state legislatures are the following. Should one be able to be certified to teach through an undergraduate pathway or is a graduate pathway necessary? Are university-based programs and low-residency programs equivalent? Can program quality be determined via analysis of program documents (course syllabi, etc.)?

*The goal of our conceptual model for physics teacher education*

In that sense, the major goal of our model is relatively simple. It needs to explain what physics teacher educators should do (and why) to prepare physics teachers who *in the fog of war—the classroom—are able to create, feed, and sustain a community of learners that engages in physics practices by deploying reliably and productively complex physics-adapted teaching methods*. The purpose of our paper is to inform the discourse in the community of physics teacher educators.[2]

## II. EXTANT RESEARCH AND MODELS FOR TEACHER PREPARATION

When one speaks of teacher learning (in the context of preparation, for instance), one typically thinks of knowledge and skills pre-service teachers (PSTs) need to develop in order to be successful when they become in-service teachers [9]. Several approaches to conceptualizing teacher knowledge are discussed in the literature [10–12]. In addition to asking what a teacher should know, we also should ask what a teacher should be able to do, *i.e.,* which teaching tasks the teacher will be required to accomplish [13]. This question offers a different perspective on teacher education in which *Tasks of Teaching* [14] are used as the starting point of discussion about teacher education [15]. We start our inquiry into teacher preparation research with the discussion of learning in general and how it applies to learning to be a teacher.

### A. Extant literature: learning to be a teacher as acquisition and as participation

Two metaphors for learning characterize the extant literature. They are significant because the design of a teacher preparation program follows from what the program leaders imagine learning to be. The two different metaphors for learning are learning as acquisition of knowledge and learning as participation in practice [16]. The former conceives knowledge as a construct of individual learners that is acquired or constructed and "resides" in an individual's brain (the cognitive, constructivist perspective), and the latter sees knowledge as "knowing" situated in real circumstances and learning as increasing participation in communities of practice (the situated, sociocultural perspective) [17–21]. The participation perspective focuses more on thinking-in-action and acting in particular circumstances and situations than on general knowledge.

However, as researchers have noted, choosing only one metaphor for learning, acquisition or participation, can leave us unprepared to deal with and better understand complex learning situations [16,22–24]. A comprehensive approach to understanding teacher education should therefore take into account both perspectives on learning.

It is also crucial to note that no matter how one conceives learning, and hence, the corresponding knowledge and skills, their productive use by the teacher is not guaranteed [25,26]. Teachers' behavior depends heavily on their often unconscious beliefs and attitudes – what we will call dispositions (see also subsection III.D.1.) – about teaching and learning [27–29]. By dispositions, we mean those beliefs and attitudes that are relatively resistant to change, and as we will further discuss, are intimately linked to teachers' habits. Dispositions influence in critical ways how teachers interpret new information, frame situations, and guide their actions [30]. Therefore, if we wish to prepare PSTs to productively apply their knowledge and skills in real classrooms, paying attention to their dispositions is paramount [12,31].

---

[2] Because of the differences between pre-service education and in-service professional development, we choose in this paper to focus exclusively on preparation.

In subsection IV.B. we define *what* knowledge, skills and dispositions well-prepared teachers should have and what tasks of teaching they should be able to perform. However, those are only one part of what teacher educators need to think about. The other part is *how* PSTs will learn what they need to know and do [9]. In the next subsection we outline the important features of teacher education programs, which work toward helping PSTs learn what they need to learn.

B. Extant literature: important features of successful programs of teacher preparation

The *general* education literature offers plenty of recommendations on running teacher preparation [32–35], induction [34,36] and professional development [24,34,37,38] programs. On the other hand, studies of *physics* teacher preparations programs also offer numerous suggestions concerning the elements of teacher preparation [39,40]. In this subsection we organize these two bodies of literature around three main ideas: *dispositions, knowledge, and skills*.[3]

- *Attending to pre-service teachers' dispositions:* Prospective teachers´ existing dispositions gained through their experiences as students influence their learning and guide their actions. Therefore instead of ignoring them, we need to address PSTs existing dispositions intentionally and critically, and further develop them in a way that will benefit their students' learning. Creating positive experience and images is an important mechanism in developing teachers' attitudes and beliefs about learning, teaching and subject matter. General education references on this matter are [26,32,34,41]. In physics teacher preparation the importance of attending to dispositions can be found in references [12,42–46].
- *Providing sufficient duration and coherence with compelling vision:* PSTs should be expected to repeatedly observe the kind of teaching that is desired of them in order to develop dispositions that are compatible with the program's vision of good teaching. The course work and clinical practice should send coherent messages to PSTs and occur over extended periods of time [33,45]. General education references on this matter are [32–34]. For physics teacher preparation references in this matter, see [12,13,44,47,48]. The program should also consistently and persistently engage PSTs in teaching practice that reinforces the vision of the program and the values of the communities in which the program is embedded.
- *Developing knowledge:* Teachers need knowledge of students and how people learn, knowledge of pedagogical methods and strategies of classroom management, and crucially, knowledge of their subject matter and domain specific pedagogical knowledge. Teachers need to know the "key modes of inquiry and thinking" and what "key ideas are foundational in their field" (p. 404) [33] and be able to contrast them to those of other disciplines. For references addressing physics-specific knowledge for teaching, see [12,13,48–50]

---

[3] In the teacher education literature, lists—even lists of powerful elements (for example [148–150])—abound. To be sure, lists of research-validated practices in which effective teachers engage can be very useful to both teachers and teacher education programs. In this paper, we seek to go beyond lists to a conceptual model that underlies such lists.

- *Providing opportunities for teaching experience:* It is of critical importance that PSTs get structured and repeated opportunities for clinical practice in order to develop practical skills and mechanisms for responding productively to constraints and opportunities of real teaching situations [26,32–35,39,51,52]. For references addressing this matter in physics teacher preparation, see [12,45,48,53,54].
- *Shaping a community:* Learning to teach means becoming a part of the community of teachers and is happening in a community of other students, teachers and teacher educators. Collaboration and support are the pillars of successful integration into cultural practices of a community. The programs should therefore aim to shape teaching and learning communities that work towards and perpetuate a shared vision of good teaching and integrate students into such communities [26,32–34]. Windschitl et al. [55] emphasize the role that tools can play, when used collaboratively, in the induction of novice teachers. Furthermore, there is extensive literature on the use of the collaborative study of student work, mostly at the in-service stage. (Critical Friends Groups [55,56] and Looking Together at Student Work [57] are examples of such approaches.) References that address the role and importance of the community in becoming a physicist and a physics teacher include [12,44,45,58,59].
- *Preparing for life-long learning:* The programs should aim to develop teachers that reflect on their practice, are prepared and willing to be life-long learners, and take an inquiry stance towards their teaching – characteristics of great importance in the fast-changing world of today [26,32–34].

The chapters in [39] address several of these points. Similarly, in his review of teacher education in physics, Meltzer writes [48] (p.11):

"[C]ertain themes have appeared in the literature with great regularity. Evidence has accumulated regarding the broad effectiveness of certain program features and types of instructional methods. The major lesson to be learned from the accumulated international experience in physics teacher education is that a specific variety of program characteristics, when well integrated, together offer the best prospects for improving the effectiveness of prospective and practicing physics teachers. This improved effectiveness, in turn, should increase teachers' ability to help their students learn physics. These program characteristics include the following:
(1) a prolonged and intensive focus on active-learning, guided-inquiry instruction; (2) use of research-based, physics-specific pedagogy, coupled with thorough study and practice of that pedagogy by prospective teachers; and (3) extensive early teaching experiences guided by physics education specialists."

The recommendations of the T-TEP Report [1], which encapsulated the findings of the Task Force on Teacher Education in Physics, reflect these program characteristics.

In both the general and physics-specific teacher education literature, therefore, there is emphasis on providing prolonged, thorough, and extensive experience to prospective teachers. This experience shapes and, reflexively, is shaped by productive knowledge, skills, and dispositions.

At this point we must note that the main goal of our framework is to go beyond finding the overlap of previous literature on the design of teacher preparation programs. This paper strives to provide a more fundamental understanding of *why* such attributes are central to successful teacher education. The current state of attempts at theory building is outlined in the next subsection.

C. Extant literature: attempts at theory building in teacher preparation

Most theoretical approaches to teacher education conceptualize it as *a process of learning to enact a certain set of high quality teaching practices.* They generally see teacher learning as becoming a part of communities of practice through practical and cognitive apprenticeship [26,32,34,51,60–66]. Perhaps the two most comprehensive approaches to teacher education from a theoretical perspective are by Ball and Cohen [26] and Hammerness et al. [32].

Ball and Cohen [26] argue that in order to learn and sustain their expertise, teachers must engage in an inquiry approach to teaching. An inquiry approach to teaching means that the teacher continuously investigates their own, and their students' thinking and actions, with the goal of improving the teaching practice. They frame *learning how to teach* as a continuing process inseparably connected to teaching practice and reflection, be it in teacher preparation programs, or throughout a teacher's professional career. Such practice-based framing of learning to teach has implications for teacher education programs.

Hammerness et al. [32] explain teacher preparation as *a process by which "new teachers learn to teach in a community that enables them to develop a* vision *for their practice; a set of* understandings *about teaching, learning, and children;* dispositions *about how to use this knowledge;* practices *that allow them to act on their intentions and beliefs; and* tools *that support their efforts"* (p. 385). The community's role, as proposed by this framework, is to enable new teachers to construct a vision of good teaching by exposing PSTs to powerful images of good practice and helping them to develop the necessary knowledge, skills, and dispositions. This can be achieved by making PSTs active members of the community of students and teachers involved in the teacher-preparation program.

Our own attempt at making a conceptual model for physics teacher preparation builds on the above-summarized research. On the basis of existing empirical and theoretical literature on teacher education in general, and physics teacher education specifically, we wish to propose a theoretical perspective that would allow a *more* fundamental understanding of the critical features of successful physics teacher education programs and provide the community of physics teacher educators with a new way of looking at and communicating about physics teacher education. We describe our proposed model in the next section.

III. DESCRIPTION OF OUR PROPOSED CONCEPTUAL MODEL OF TEACHER PREPARATION

In this section we describe our proposed conceptual model for teacher preparation that will allow us to explain the existing recommendations from the previous section and to guide the design and improvement of teacher preparation programs. Although the conceptual model is generic, in the sense that it is formulated for all content areas, its usefulness for teacher preparation programs is manifested *only* if its components are articulated in a particular knowledge domain (physics, mathematics, history, etc.)

If we examine the empirically derived bulleted list in subsection II.B and the theoretical discussion in II.2 C we see the unifying theme of continuance and coherence. Teachers need to continuously inquire into their own learning and practice and continuously participate in communities of practice. The necessity of this immersive type of continuance emerges as the very fundamental building block of teacher preparation. Why is this the case? We need to prepare a teacher who in the challenging and unpredictable circumstances of real-time school lessons will implement high quality instruction – no matter what. This is more likely to happen if instructional responses to continuously varying instructional stimuli are habitual. Thus we hypothesize that the foundation of teacher preparation that explains the existing features of teacher preparation programs is that all good programs ultimately try to form *habits, namely what a teacher thinks/does spontaneously, especially under the situational constraints of everyday circumstances of teaching practice.* [67] As the ecology of needed habits is rich and complex and each habit takes time to develop, the need for *continuance* arises. This way the need for habit formation explains the continuous nature of teacher development that unifies empirically found features of successful teacher preparation programs.

Extant literature on teacher preparation in many ways touches on the concept of habits They are often conceptualized as being integral parts of belief systems [30,32,61,68]. Dewey sees habits as being more than just unconscious automated behaviors cemented by repetition. In fact, he suggests that they are intimately connected to our dispositions (as we have operationally defined them in II.A) [69,70] and places them at the very core of human physical and social function.

We argue that many of the features of existing theoretical frameworks and descriptions of best practices can be *better* understood if we recognize the centrality of habits in the everyday work of teachers and the importance of productive habit development in the preparation of new teachers. The importance of habits is something for which Dewey advocated almost a century ago. By situating habits as the cornerstone of our conceptual model, we hope that novel insights will flourish and new perspectives will ultimately take over.

The main idea of the model that we propose is that the teacher preparation program needs to develop desired habits that are based on knowledge, skills and dispositions, whose development is nurtured by a community. This interaction of knowledge, skills and dispositions within the community of like-minded teachers feeds habit formation.[4] The following subsections provide the assumptions and overview of the model, and instantiate the model in the physics domain.

A. Assumptions

Before we describe the model we list assumptions that we take as givens. We have grouped the assumptions into larger groups. Most of those are general – they apply to any field, but some are physics specific.

*Assumptions about what constitutes learning*

Assumption 1: People learn by engaging in social activities that allow them to construct knowledge together in interacting with appropriate tools and each other [19,20,44,71].

Assumption 2: A primary goal of education in a high school is to enculturate high school students in the practices of the discipline through learning a limited number of fundamental ideas (in physics, these can be conservation laws) and crosscutting concepts that transcend all sciences (such as systems and patterns) [72]. In physics such enculturation involves learning to use intuition and creativity to induce physical principles from observational data; to devise and test coherent mechanistic and mathematical models of observed physical phenomena, which at first seem incomprehensible and mysterious; to test and revise ones' ideas about the physical world through a rigorous practice of experiment and analysis; to experience the joy of experiment and discovery through hands-on practice; to learn to base one's thinking on experiment and reasoning, rather than solely on authority.

---

[4] (Mis)appropriating John Archibald Wheeler's quote, "Spacetime tells matter how to move; matter tells spacetime how to curve," we claim that the community shows PSTs what habits should be developed and PSTs' emerging knowledge, skills, and dispositions are the prime materials from which the community forms these habits.

Assumption 3: Apprenticeship is one of the most effective models of learning. Historically this mode was used to learn any craft (painting, cooking, sewing, etc.) Recently learning theories started including cognitive apprenticeship [60,73], which is different from traditional apprenticeship, since the skills of the master are invisible and certain "unpacking" and reflection are needed. The apprenticeship approach is especially relevant for physics where physics researchers learn most of the research skills through apprenticeship, not through formal coursework (we need a ref to how one becomes a physicist [18–20].

Assumption 4: Before they start the program, PSTs have strong expectations, based on their own experiences and cultural images, of what physics learning entails and therefore what "good" teaching is [68].

*Assumptions about teacher education programs*

Assumption 5: Discipline-based education research generates reproducible results that should guide how we approach teaching of a discipline and consequently how we prepare teachers.

Assumption 6: We assume that a discipline-based teacher education program also provides opportunities for future teachers to study general pedagogical issues: psychology, diversity, motivation, group work, etc. We also assume that the program staff makes an effort to connect general education courses to physics-specific education and clinical practice.

In addition to articulating the previous assumptions, we list also the following limitations of our approach. In this paper we do not discuss the many factors contributing to recruitment of future teachers or issues associated with professional development and retention of in-service teachers.

B. Overview and elements of the model

In this subsection we discuss the elements of the model and their interrelationships. In sections IV and V we show how this model is applied in the context of physics teacher preparation.

*Desired outcome of a teacher preparation program: cultivation of appropriate habits*

At the end of the program we envision a member of the education community in a specific knowledge domain (physics, mathematics, English, etc.) who has developed, in addition to the requisite knowledge, skills and dispositions, the *habits* of learning and teaching that are consistent with our current knowledge of what processes in learners lead to effective learning and what teachers should do to promote learning [74].

*Role of habits in psychology*

Research in psychology points to the role of habits in bridging the gap between attitude/intention and behavior [75] and their role as a "repository of chronic goals and motives" [76]. Studies of the relationship between habitual and thoughtful decision-making suggest that habitual responses generally invoke less conscious consideration of the situational requirements than intentional behavior [77]. Therefore, a question may arise whether deliberate development of habits in PSTs is not doing the exact opposite of shaping a teacher capable of managing complex situations. We argue that this is not necessarily the case. Automated behavior is not necessarily simple and linear and even though it is habitual, it can at the same time be responsive to situational requirements [76]. Korthagen & Lagerwerf [27] suggest that *Gestalts*, precompiled conglomerates of unconscious sources of behavior, such as feelings, routines and values, are activated in response to situational cues, and take central part in shaping teachers' behavior.

*Productive vs. unproductive habits*

Teachers unavoidably develop habits with practical experience and under the influence of knowledge and belief structures that in many ways condition the responses of teachers in their practical work. This way, many of the tasks of teaching become more automated with time, which also allows teachers to devote more attention to other tasks. However, the habits that teachers develop when they enter teaching for the first time will often be shaped by the pressure and demands of a very dynamic environment and may as a result be directed towards teacher "survival" instead of student learning. It is therefore very important that the teacher "bring with her" a set of habits that will allow her to get a grasp of the situation and steer her practice in a direction that will lead to student learning, as well as her own advancement of teaching. Thus, the habits we have in mind are not meant to be enacted as inflexible, mindless scripts for teaching, but rather as a toolkit or a support structure of strategies for automated behavior that will allow a new teacher to successfully start practicing reformed teaching and help her avoid picking up unproductive habits under the pressure of real teaching environments or reverting to the "old ways", learnt through the apprenticeship of observation, which are in most cases less supportive of student learning.[5]

Our model addresses the preparation of teachers for *routine decision making* [78,79] and brings it to the center of attention by regarding teachers' habits as an integral part of teaching practice and their formation as an essential part of learning to teach [69]. These habits include *domain specific content* and *teaching* habits of mind (reasoning like an expert in a variety of situations related to learning), habits of *practice* (making and enacting decisions during lesson preparation and in-class instruction that are consistent with promoting student learning, and reflecting on one's actions and student learning after each lesson) and habits of *maintenance and improvement* (continuously engaging in personal and communal reflection and professional development, and exercising multivalent leadership roles to improve habits of mind and ensure the general wellbeing of all students). How can these habits be long-lasting and generalizable beyond very specific contexts? It is critical that the automatic behaviors are acquired through conscious and deliberate focus directed towards an ideal concept of optimum practice [80,81].

To emerge from the program with such habits pre-service teachers need to develop necessary knowledge, skills and dispositions. These can only develop within a community of like-minded people who are working toward the same goal [17] and with similar vision of what constitutes rich learning for all students.

*DHAC model: analogical and graphical representations*

Figure 1 represents our proposed model for a prepared teacher, as well as the process that shapes a candidate in a teacher preparation program. To help us convey the key aspects and processes of teacher preparation we use an illustrative analogy of a tree (the teacher) growing in the environment (the community of the teacher preparation program). In this analogy, the parts of the tree illustrate our conceptualization of the characteristics of a prepared teacher. The process of shaping the teacher is illustrated by the interaction of the tree with the environment.

---

[5] Astronauts are trained to develop habitual responses to challenges, with full knowledge that if an emergency arises in space it will most certainly be in a context that was not fully anticipated.

In the analogy, the roots represent the grounding of the teacher in the knowledge, skills and dispositions lived out by the community of the program. The trunk of the tree represents the teacher's habits of mind and practice, as they have been shaped by his or her interaction with the communities of practice. The canopy represents the teacher's capability to grow and improve professionally, which manifests itself in our model as the habits of maintenance. It is through photosynthesis in the canopy that the tree grows, which metaphorically illustrates the process of professional growth of a teacher. Note that the parts of the tree make up a wholesome organism and grow in concert, affecting each other's development. The analogy helps us illustrate that all components that make up the model of the teacher are essential in accounting for a successfully functioning and continuously developing teacher. In the model, the habits of mind and habits of practice (represented by the tree trunk) enable the teacher to "stand on their own feet" in the unpredictable circumstances of the teaching practice and to draw productively on their expertise (the roots). A teacher must be able to weather the pressures of everyday demands, after they leave the program. This requires emotional and professional sturdiness, embodied in the habits of mind and practice, as illustrated by the tree trunk.

The complementary part of the model (the environment in the tree-environment analogy) stands for the process of the preparation of teachers. The environment in which the tree grows represents the communities of practice that nurture the teacher as he or she progresses through the program. If the teacher is to be properly prepared to teach in real circumstances, the program must structure the PST's involvement with the communities of practice so that he or she develops the requisite habits. In the analogy, this means that the tree should be exposed to sufficient stress to grow a strong trunk, but should at the same time be exposed to positive experiences of teaching (sunlight) and nutrient-rich water to develop healthy roots that serve as a grounding in the knowledge, skills and dispositions perpetuated by the community. We hypothesize that this process requires a coordinated, sustained and coherent effort of the community, not only because of the time required to develop the "roots", but equally importantly, the time and effort required to properly develop the "trunk" and the "canopy". Going along with the analogy, the shaping of habits is a process that requires patience, firm scaffolding, and tools [55,82,83].

The tree-environment analogy helps us convey the central aspects of our model for teacher preparation on a structural level. If the model is to become useful for informing the discourse in physics teacher preparation, it needs to be "furnished" on a finer level. This means that we need to substantiate what habits of mind, practice and maintenance, as well as knowledge, skills and dispositions are important for the preparation of *physics* teachers and how they can be developed. We do that in section IV. First, however, we define what we mean by these terms.

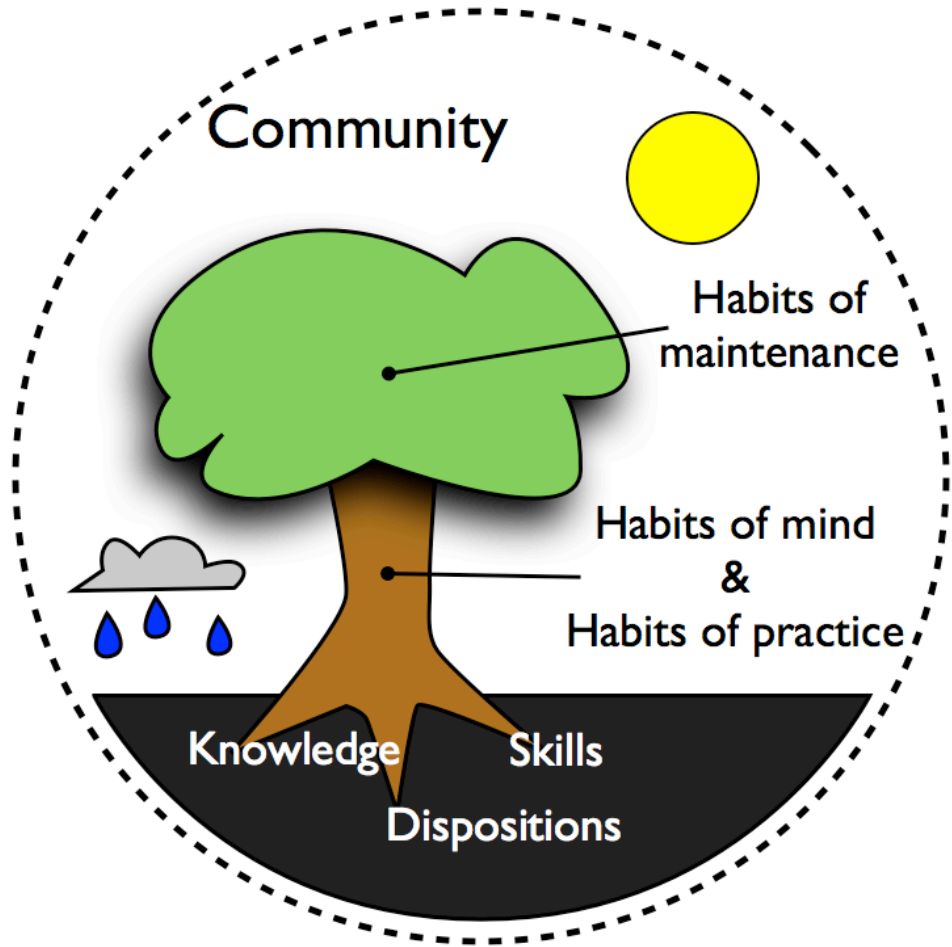

Figure 1: (Color) An illustrative analogy of the DHAC conceptual model. The teacher (the tree) is shaped by the community (the environment) in which it develops.

*Knowledge, skills, and dispositions: definitions for DHAC model*

*Knowledge:* Much has been written about the nature of teacher knowledge and the differences between content knowledge, pedagogical content knowledge, and content knowledge for teaching a particular subject. The readers can find relevant information in [10,84–86]. In subsection IV.B.2 we focus on the knowledge that is required to develop the habits of mind and practice of a physics teacher.

*Skills:* For our purposes, a skill is a pre-compiled procedure that one deploys automatically without consciously thinking about it. We break down these skills into mental, technical, and emotional skills. The mental skills relate to operations that a teacher needs to execute on a regular basis and in the spur of the moment. For example, reflecting critically on one's lesson is a skill that a teacher needs to use after each lesson while changing the flow of a lesson based on a student comment is a skill that is executed at the "right" moment. Technical skills relate to the ability to operate and fix equipment and to operate modern technology. Technical skills are especially important at smaller schools where there are no lab assistants or in schools where resources for equipment are scarce [87], for example). Emotional skills relate to the ability to keep cool no matter what happens in the classroom and the ability to put children's welfare and learning first. Although there seems to be a thin line between a skill and a habit as both are automatic and often subconscious we treat skills separately because being skilled in something does not necessarily mean doing it habitually. For example we are all skilled in drinking water but not all of us have a habit of drinking the requisite number of glasses a day.

*Dispositions:* In the dictionary "disposition" is defined as "a person's inherent qualities of mind and character" [88]. Such definition does not fit our meaning of this word in a teacher preparation context. We define a disposition as a strong (often subconscious) belief or attitude related to some aspect of teaching. Imagine a PST who believes deeply that physics is a difficult subject and only a few selected students can learn it. How does this belief affect the choices that this teacher will make in the classroom? An example here would be assessment. If the teacher believes that only the brightest students can master physics, then the role of assessment may be interpreted as a one-time grading filter that distinguishes more and less able students. If on the contrary, the teacher believes that all students can learn physics but need different amount of time and effort, then this teacher will set up a system where a student can continuously improve her work and consequently, improve her grades. As we can see from this example, the difference in the belief of student ability to learn leads to huge differences in classroom instruction.

Taking PSTs' dispositions into account, learning to teach begins long before PSTs enter teacher preparation programs. Out of all kinds of professional learning, learning to teach is special in the sense that all future teachers have, in the course of their own education, observed other teachers working for thousands of hours before they themselves started formal teacher training. This so-called "apprenticeship by observation" [89] is usually considered to be more potent than formal teacher education [9,26,32,34,68]. The robustness of PSTs' existing dispositions pose a challenge to teacher educators since they play an important role in how teachers perceive new information, frame situations and act in practice [30,90]. Therefore, teacher educators should not ignore the central role of PSTs' attitudes and beliefs in the process of learning to teach [91]. Good teacher education programs should, besides giving attention to prospective teachers' knowledge and skills, aim to develop their dispositions in ways that will be conducive to reformed[6] teaching [34]. However, changing people's beliefs may be a challenging and long lasting process [68,92].

*Need to "furnish" the model*

So far, we have presented the structural outline of our model. At this level, it can still be used in different subject domains. However, because different disciplines bring with them diverse ways of understanding and seeing the world, the model needs to be "equipped" or "furnished" to suit specific subject-domains and their "signature pedagogies" [93], if it is to be useful to inform the preparation of teachers in those disciplines. In the next section, we "furnish" the model to accommodate physics teacher preparation. In other words, we put in place some of the content of the categories, which we have so far discussed on the structural level. This includes explicating some physics-specific components of the model (*e.g.* physics-specific habits of mind), as well as some more general components (e.g. the habit of listening to student ideas), which are nevertheless essential for successful physics teaching. Such "furnishing" is based on the analysis of the literature and our experience. Time and further research will show how useful our ideas are for practical instantiation of effective physics teacher preparation programs.

---

[6] By reformed we mean teaching that engages students in an active contruction of their own ideas through participation in physics-rich group work.

IV. USING THE MODEL TO CONCEPUALIZE A PREPARED PHYSICS TEACHER

In this section, we apply our proposed model to the special case of physics teacher education. In subsection A we outline productive habits of mind, of practice, of maintenance, and of leadership, both physics-related and physics-teaching related, that a physics teacher education needs to inculcate in PSTs. In subsection B, we outline the knowledge, skills, and dispositions that undergird such habits. We also include some habits that are not exclusively physics-specific but are crucial for a physics teacher to develop.

A. Habits of a physics teacher

Below, we outline the specific habits of mind and habits of practice that need to be developed by a prospective physics teacher. Consequently, when one designs/reforms a teacher preparation program, it is good to keep in mind the development of the following habits.

*Habits of mind:* these are the examples of habits of thinking like a physicist and thinking like a physics teacher. The list below was compiled based on the history of physics, physics education literature and observations of the work of physicists in real time [94–97]. We recognize that the list below is not exhaustive but we hope that it will give the reader a flavor of what we have in mind. By physics habits of mind we mean spontaneous thinking of and noticing the application of physics ideas in the surrounding world and in the "playground" of other disciplines, such as mathematics, specifically discerning the most important factors affecting the behavior of a certain system and modeling said system in terms of a simple model. By physics teacher habits of mind we mean spontaneous thinking and attending to student physics-related reasoning, questioning, and development.

*Habits of practice* include (a) the habits that involve spontaneous decisions during lesson planning and (b) the habits, which enacted in the classroom, lead to student learning. The habits of practice are therefore intertwined with the habits of mind and cannot be separated definitively.

Finally, *habits of maintenance and improvement* are the habits that involve continuous learning on the part of the teacher as an individual and as a member of the community, as she organizes her professional life to give priority to maintaining the community, actively sharing new findings and using the findings of other teachers.

Selected *physics habits of mind* include:

- Seeing physics everywhere (for example, thinking of static friction when walking, noticing diffraction pattern on eye lashes when squinting, etc.)
- Approaching problem solving as a physicist (napkin calculations, drawing a sketch before solving any problem, being able to do an order of magnitude estimation, being able to do a long calculation without a calculator just using powers of 10, etc.) [98,99]
- Treating physics as a process, not a set of rules or a collection of information, specifically seeking to understand how physics ideas have emerged in specific historical contexts, how they are connected with progenitor ideas and how they have given rise to subsequent ideas [93,100]. This habit of mind protects against naive hero worship by providing more nuanced understanding of the ecology of physics concepts as invented intellectual entities. Examples of such epistemological habits of mind include inductive (experiment-based) and "spherical cow" reasoning, analogical reasoning, establishing causality, questioning claims, quickly assessing coherence of suggested ideas with the rest of the physics body of knowledge, and being able to spontaneously think of an experiment to test an idea when it is proposed (hypothetico-deductive reasoning).
- Using mathematics in a physics-specific way [101]. Specifically, mathematics plays a different role in physics compared to other sciences. Physics is much less statistics-oriented than biology and more oriented towards mathematical modeling and internal consistency of multiple representations [99] than chemistry.

Selected *physics teacher habits of mind and practice* include:[7]
- When helping students learn, starting by helping them connect new ideas to their existing ideas recognizing that student ideas about a specific set of physical phenomena have been developed over years of organizing, perhaps uncritically, a rich set of experiences, full of complexity and messiness. School physics, on the other hand, is often sanitized and deals with idealized contexts. In the same way that to Greek two-year olds, their language is not "Greek to them," students can be extremely successful in predicting, with almost zero conscious analysis, where a tossed ball will land or how to place boxes of different sizes on each other so that they won't tip over.
- Attending to students' thinking specifically regarding physical mechanism, interaction, conservation, constraint, and processes of change and transformation.
- Encouraging students to test their ideas experimentally instead of waiting for validation from authority. For example, in physics, it is often easy and highly productive to say, "Go ahead and try it; set it up [bulbs, cart, magnet] and see what happens." In chemistry, that may not be true, and in biology, it is very unlikely to be true; that is, you would need to make a different "spontaneous decision."
- Listening to student conversations, comments, and questions related to physics (both content and practice) and altering, revising, improvising planned instruction to build on students' ideas.
- Seeking to establish a step-by-step causal chain to understand a particular phenomenon or to search for a mechanism (or a set of mechanisms) instead of just describing the phenomena by collecting and analyzing data.
- Reflecting on the role that language plays in student learning and making conscious choices of words and grammatical construction when talking about physics so as not to create confusion (for example, knowing that "heat" means different things in physics and in the daily life of students; or choosing the appropriate time to differentiate gravitational potential energy in an object from the energy of a system of gravitating objects, etc.)
- Treating all students as capable of learning physics and contributing to the generation of physics knowledge (as opposed to treating learning physics as a weed-out competition).
- Being aware of the "surroundings" (nature, current events, such as breakthroughs in science or

---

[7] We assume that generic habits of practice such as preparing every lesson carefully while attending to three components: learning goals for students (what will students learn), assessment (how will the teacher know that they learned) and activities (what will students do to learn) are also being formed but as they are not physics specific, we are not focusing on them here. The same is true for the habits of reflection.

socio-scientific issues, such as climate change, etc.) as a source of learning physics (for example, watching waves going over a rock in the sea and taking a video to use in the lesson on diffraction) by building on the inherent ease of experimentation that physics affords; habitually thinking of how to use an everyday simple phenomenon, video, etc. to help students notice, wonder about, and learn something (for example, stumbling upon a video on YouTube and immediately incorporating it in the lesson the following day); dumpster diving (not passing by something in a dumpster that can be used for helping students learn physics) through a conviction that cheap, readily-available materials can serve as the basis of a good physics lesson.

None of these habits can be developed without deep connections to physics as an experimental science and to physics education research, which provides PSTs with tools that help build these habits. For example, what are productive questions to ask students who are learning electric current to make sure that they understand the difference between series and parallel connections? How do you make sure that the Wimshurst generator that the school owns lasts for a long time?

Finally, *habits of maintenance and improvement* need to be developed. For example, being a member of the AAPT and habitually reading The Physics Teacher is one of those habits. Being a member of the local section of the AAPT, attending workshops led by local teachers, or being connected to the physics department of a university are additional examples of such habits.

An important subset of these habits consists of *habits of leadership.* Habits of leadership involve spontaneous steps to engage other teachers in thinking about student learning and showing examples of how to do it. For example, having a discussion about the implementation of NGSS in the teachers' lounge during lunchtime, sharing new ideas learned in a workshop with other teachers, etc.

We have operationalized the above habits in the context of physics teacher education. As Meltzer puts it, "The[se habits] should be integrally linked to the emotions experienced by physicists in wondering about, exploring, and discovering underlying coherence in the world around us; else they may become mere boring drudgery, analogous to adding columns of numbers. Teaching habits should be linked to effective engagement in joyful communication of new ideas and abilities to interested students; the alternative is obvious, and depressing to contemplate." [80]

### B. Dispositions, knowledge, and skills

In this subsection we describe dispositions, knowledge, and skills that form the foundation for the habits of mind and practice described above and follow from our assumptions described in subsection III.B. We seek to have a minimal set of inputs in our model. All the inputs have to either be founded on our assumptions or serve the grand unifying theme. We start with dispositions as they determine what knowledge and what skills the PSTs "choose" to appropriate during the program [30,68,78,91,92,102,103].

#### 1. Dispositions

Below we list the most important dispositions that we consider crucial for a successful physics teacher. This list does not attempt to be exhaustive. Although some of the listed dispositions may be desirable for teachers in subjects other than physics, their impact on teaching physics is particularly critical because of the unique opportunities that physics affords for the development of students' critical thinking skills and scientific competencies, which themselves arise from the nature of physics knowledge and the process of the development of this knowledge.

- Learning of physics is *doing* physics. Just as physics is a specific process of inquiry into nature and not a static set of rules to be mastered, learning physics is a process that should reflect the nature of physics as both an experimental and a theoretical science[8]. Learning of physics involves learning of the process of physics as much as its final outcome [100,104–106].
- Learning of physics is a social co-construction of knowledge through experimentation and reasoning that leads to changes in the brain of the learner [107]. The roles of the teacher in this process is to guide, provide informed feedback, and support [108].
- Learning is a complex process, which cannot be reduced to the following dichotomy: student either got *it* or did not get *it*. The a-priori position of the teacher should be that all students are capable of learning physics, but every person learns at her/his own pace. Research shows that teacher expectations can influence student achievement [109].
- Learning of physics does not happen in a vacuum—everything that happens in the classroom is affected by a student's experience, language and environment. [25,74,110]
- Intrinsic motivation is to be treated as the crucial attendant to learning. In physics, student motivation comes from noticing that one is becoming more successful; from being able to apply physics knowledge to relevant everyday experiences; and from experiencing the inherent intellectual excitement of physics [111–114].
- All decisions in the classroom are to be guided by the "learning compass." If a certain decision is likely to lead to more learning, it is a productive decision; if it is likely to impede learning, it is counterproductive. These decisions are often triggered by students' comments or questions that require the teacher to alter the existing lesson plan. The readiness to change the plan using the above compass is the disposition of a reformed teacher [25].
- A teacher is always a learner and aspires to become an expert learner.

The dispositions described above might serve as a filter for the knowledge that a PST appropriates during the program. The above dispositions are to be developed throughout the program. Note that although the dispositions might seem generic, they are intrinsically physics specific. For example, a common view of physicists in society is that physics is only for very smart people and thus not everyone can learn physics. Sharing the disposition that all students can learn physics and developing habits of practice built on this disposition will dramatically change the number of students successful in physics and the image of a physicist in the society.

### 2.    Knowledge

To develop the physicist's habits of mind, a program needs to foster the following knowledge.

- *Habit: noticing physics in the surrounding world.* One must have a very deep knowledge of physics as a body of knowledge (at the level of and beyond a general physics course), and this knowledge should be both at a conceptual and quantitative level. The conceptual level involves the ability to talk about any physical phenomenon, mechanism or relation using simple words and images (these include different physics-like representations such as force diagrams, sketches, ray diagrams, field diagrams, etc.) that non-physicists understand [115]. The quantitative level involves the ability to use appropriate mathematics (this means finding the best simplified model) to analyze a relevant phenomenon or a process and to evaluate the result of a calculation. For example, watching a YouTube video of some sensational experiment, a person with physics knowledge should be able to analyze it using appropriate tools to decide whether it is a real one or a fake.

---

[8] The recent national attention to computational literacy lends itself to an additional focus of physics instruction, namely physics as a computational science.

- *Habit: approaching problem solving as a physicist*. Physicists do not solve problems by searching for the right formula to plug in the givens (the novice approach) [99]. Instead they draw a picture of the situation, think of it conceptually first without searching for equations, appeal to grand principles rather than superficial features, solve equations symbolically before plugging in any numbers, and finally evaluate the solution. Traditionally, the types of "problems" (more appropriately, exercises) that students encounter in general physics courses tend to be short, quick applications of one or two formulas [116]. At that stage of study, many of the students do not know that they will become physics teachers. Later, when they start working towards their physics education degree, they are usually enrolled in upper division physics courses that do not involve problem solving that is as relevant to their vocation. This leaves PSTs at the level of novices with respect to the high school level physics problems. Thus in the teacher preparation program special attention should be given to the procedural aspect of problem solving of typical and atypical introductory physics problems so the PSTs develop expert-like problem solving strategies [95,96].

- *Habit: Treating physics as a process, not a set of rules*. To develop this habit the PSTs have to know how the canonical knowledge came to be and to appreciate how physicists developed and discarded ideas. This knowledge will enable PSTs to make progress developing sophisticated physics epistemology. This knowledge should be combined with the knowledge of student development of the same ideas. Historical knowledge examples include the development of understanding of concepts of velocity and acceleration (Galileo), force (Newton), momentum (Descartes, Huygens, Leibnitz), energy (Mayer, Joule, Helmholtz), electromagnetic field (Faraday, Ampère, Maxwell, Hertz), light (Newton, Young, Planck, Einstein), atom (Rutherford, Bohr), nucleus (the Curies, Meitner), and so forth. In addition, to develop this more sophisticated physics epistemology PSTs need to have some research experience, be it in physics or in physics education research.

- *Habit: Using mathematics in a physics-specific way*. To develop this habit the PSTs need deep conceptual and not just procedural understanding of the nature of physics equations and the difference between operational definitions (density is equal to the mass divided by volume) and cause effect relationships (acceleration is equal to the sum of the forces divided by mass); functional understanding of proportional reasoning, the difference between the meanings of math operations in mathematics (where multiplication can be always reduced to addition) and in physics (where it often cannot be, for example $p=mv$). The work of Bruce Sherin, Joe Redish and colleagues, Suzanne Brahmia, Steve Kanim, and many others can help physics teacher educators develop appropriate habits of mind [101,117–120].

The next step is to discuss the knowledge that is necessary to develop habits of thinking like a reformed physics teacher and habits of practice. Here by reformed we mean a teacher who believes that the role of a teacher is to create an environment for the students to learn, not to provide the best explanations of observed phenomena as possible.

- *Habit: Helping students connect new ideas to their existing ideas and applying them to real-world phenomena*. Here the knowledge of students' ideas is invaluable. It has been documented in the literature and PSTs definitely need to be familiar with this work (See [121,122] for exhaustive lists of references). However, this knowledge by itself is not enough. Knowing student ideas should lead to the knowledge of how to anticipate student thinking around these physics ideas, how to promote their interest, how to monitor student development (what types of questions to ask in a specific situation (What do you mean by the word "force"?), what types of questions are unlikely to yield productive insights into student thinking in a specific context (What is energy?), how to interpret and act on student thinking (for instance, when a student says that when a ball thrown upward slows down to a stop it is because the force given to it by the hand runs out, what might the student's cognitive or experiential need be? [110]). In other words, a reform-oriented physics teacher needs to know how to take student ideas and help the students move forward in a way that is consistent with physics practice. This knowledge rests on the deep knowledge of physics as a process and deep knowledge of PER as a field of study of student learning of physics.

- *Habit: Preparing every lesson carefully attending to three components: goals, assessment, and activities*. To develop this habit the PSTs need generic knowledge of instructional planning (units and lessons), but they also need to know what this means for specific lessons. Therefore they need to be familiar with the documents discussing the goals and assessment of physics instruction (NGSS, etc.) and with the resources for physics activities and assessments (for example [123–130]). In addition, the PSTs should have strong knowledge of curriculum as a map that allows students to build new ideas on something that they already know. For example, PSTs need to be able to explain why one needs to be familiar with the kinetic molecular theory to understand sound waves and why one needs to understand the concept of a system to learn energy. This is especially relevant given research results that show that the quality of new teachers is determined by the amount of experience they have with day-by-day implementation of instructional units in their preparation programs [131].

- *Habit: Listening to student comments, and altering, revising, improvising planned instruction to build on students' ideas*. Once the lesson plan is written it is very difficult to change it "on the fly" but if the change is not made at a particular moment, a learning opportunity will be missed. What knowledge should PSTs possess in order to be able to "let go" of the lesson plan at an appropriate moment? First, they need to know why such change is important and how it contributes to learning [132]. Second, they need to be able to hear what a student is saying (this is a skill and we will discuss it in the skills subsection III.D.3.) and to know their current lesson and the lessons that will follow so well that they can see how to change the lesson in response to the comment (this goes back to the knowledge of physics curriculum). From this argument follows that the PSTs should think of planning of several lessons (a unit) not just one lesson.

- *Habit: Seeking a step-by-step causal or mechanistic understanding*. Although statistical, covering-law, etc., explanations have their place in physics, the direct interplay between cause and effect is a special strength of physics learning. One needs to know, therefore, what counts as a physics explanation of a phenomenon. For example: when a puddle of water dries, a physicist would explain it as a selective escape of the fastest moving particles while a lay person might say that water tends to turn into gas.

- *Habit: Language awareness*. A great deal of knowledge is needed here. PSTs need to develop an understanding of ways in which learners use analogical and metaphorical language and recognize the instructional affordances of the use (or misuse) of metaphorical language by students. PSTs also need to understand student difficulties with specific terms (force, flux, heat, weight) that have a different meaning in everyday life compared to physics, and grammatical constructions that change the physics meaning of a concept *(e.g.,* an object in a potential well) (see [110,133,134]).

- *Habit: Treating all students as capable of learning physics and contributing*. To do this, the teacher needs to know how to engage students of different backgrounds and levels of preparation in the learning of physics in the same classroom. For example, using concrete representations such a motion and force diagrams and bar charts with ELL students, tailoring the content of the problems to the vocabulary and experience of a specific student population, etc.

- *Habit: Being aware of the surroundings (nature, current events or societal issues, etc.) as a source of learning physics*. To develop this habit PSTs need deep knowledge of physics, knowledge of curriculum goals and knowledge of equipment.

*3.     Skills*

In section III.B we defined what we mean by skills and we described different kinds of skills that a physics teacher needs to develop. Although there seems to be a thin line between a skill and a habit as both are automatic and often subconscious we treat skills separately because being skilled in something does not necessarily mean doing it habitually. For example we are all skilled in drinking water but not all of us have a habit of drinking the requisite number of glasses a day.

Examples of mental skills include: interpreting skillfully what students say using physics language, expressing a complex physics idea in simple words and/or without mathematics; doing power of ten estimations in one's head (including remembering the order of magnitude of important physical constants); fluently using physics-related mathematical skills such as ranking possible effects according to their size or proportional reasoning.

Examples of technical skills include: Using all traditional equipment for physics curriculum; building and troubleshooting electric circuits, soldering, repairing simple equipment, maintaining equipment; having deep knowledge of document authoring and spreadsheet applications (like mail merges, graphing, etc.); using educational technology for communication with students; using physics technology for data collection and analysis; choosing and purchasing quality equipment efficiently.

Examples of emotional skills include: Empathetic appreciation of the challenges that a novice learner faces when confronted with difficult ideas or a parent faces when confronted with an assessment of learning challenges of the child; interpreting adolescent student behavior not as an immediate reflection of the teacher's popularity but as a form of expression of the student's internal state; being able to think and multitask under stress.

V. USING THE MODEL TO PROPOSE MECHANISMS FOR THE FORMATION OF HABITS

In the first subsection we discuss the mechanisms through which physics teacher preparation programs can help future physics teachers develop the knowledge, skills, dispositions described above, and the conditions that should be met to use those for successful habit development. Starting from the perspective that learners of all kinds bring a rich fund of ideas based on years and years of active and passive sense-making, PSTs bring to the program their own views of what constitutes physics learning and good teaching. Unfortunately, often these views are based on the experience with traditional instruction [135]. Thus if we wish that our future teachers develop the habits described above, we propose that a program needs to integrate three components: (a) extensive apprenticeship-based clinical practice, (b) in-depth coursework on the learning and teaching of high school physics, and (c) the care and feeding of a rich community of practice [136].[9] Our recommendations are simultaneously general and physics-specific. In the second subsection we propose a set of actions that a physics teacher education program can take right away to start on the path of incorporating our recommendations into its design.

A. Three mechanisms for productive habit formation

We propose three mechanisms for productive habit formation: apprenticeship-based clinical practice, coursework on the learning and teaching of physics, and physics teaching community of practice. We also explain how these mechanisms develop the requisite knowledge, skills, and dispositions that fund the productive habits we are after.

1. Apprenticeship-based clinical practice

Any apprenticeship begins with students observing a master performing the craft. In our case, observing reformed physics teaching and being in the role of a student in this environment is crucial. PSTs need to have images of physics learning and teaching before they can start practicing such teaching themselves. Thus the PST first needs to observe reformed physics instruction and then slowly start participating in it serving as an apprentice to the master teachers, with the master teachers providing scaffolding for the PST's tasks, guidance at the micro level of their implementation and substantive feedback in real-time [39]. This construal of clinical practice takes on urgent importance in the context of the national landscape of physics teacher education, according to which the majority of physics teachers (as opposed to biology teachers, for instance) are (a) the only such teachers in their schools and (b) do not have a physics major or minor. Placing PSTs in state-mandated teaching practica is treated often, therefore, as a matter of logistical matchmaking rather than intentional modeling of the vision of the physics teacher education program. One way to fix this conundrum is to maintain close contact with program graduates and continuously interact with them professionally so they grow into the master teachers for new PSTs [45].

---

[9] To be sure, there need to be additional components to the program that attend to important general aspects of teaching and learning, as well as to treatment of the student as a developing human being, to understanding contextual characteristics of schools and the teaching profession, to gaining familiarity with state requirements, etc.

Since habits take a long time to develop and become cemented [137], it stands to reason that the program needs to provide early and multiple opportunities for reflective practice conducted *in service* of habit development. We envision a progression of such clinical practice: from observation, reflection and analysis of specific aspects of effective teaching modeled by master physics teacher(s) (for example, a Teacher in Residence [138]) to opportunities provided to the apprentices to try out short, discrete tasks of teaching [45] in the context of a reform-centered physics classroom (university or K-12) with subsequent reflection, to increasingly more formal teaching occasions all the way through to the ultimate relinquishing of scaffolds at the very end of student teaching, as the teacher candidate has shown repeated evidence that effective habits have been appropriated. All these experiences come with an expectation of repeated attempts by the apprentice, performed in the presence of the community of the master teacher and other apprentices, and the incorporation of the immediate feedback of the master teacher. It is crucial that all "masters" participating in the process explicitly share the same vision for good physics teaching and the same dispositions as described above. (For examples of structuring clinical practice using the community-based apprenticeship, see [39].)

In learning how to teach physics the crucial element is participating in designing, teaching, analyzing, and reflecting on instruction that places emphasis on students' scientific inquiry. Many teachers have not experienced this kind of instruction and do not practice it, thus placing PSTs in the classrooms of those who indeed practice effective inquiry-based instruction is a challenge. Consequently, cooperating teachers need to be skilled in using physics experiments as a means of helping students to construct, test, and apply knowledge. In such classrooms PSTs can develop a repertoire of productive experiments, become familiar with common equipment used in schools and have an opportunity to learn how to repair equipment and how to invent new hands-on ways to learn physics. A related challenge is finding classrooms in which physics learning does not just consist of solving mathematical word problems but are an environment in which students have multiple opportunities to describe real world phenomena through multiple representations.

According to our model then, the typical generic classroom observations that are intended to familiarize the PST with school contexts, followed by the standard student teaching experience, understood here as the placement of an individual teacher candidate in an arbitrary collaborating teacher's classroom to fulfill a state statute that has a specified one-size-fits-all duration (and that often—but not always—is held in parallel with enrollment in a general science methods course that emphasizes the development, implementation, and assessment of a single lesson in great detail or at most a small unit), although valuable, are woefully inadequate in helping novices to develop a toolbox of thoughtful habits and a beginning curricular repertoire for the whole year. In our experience, it takes more than a year of this closely supervised apprenticeship-based practice that allows a personalized schedule of slow removal of scaffolding, for a teacher to be able to start the first day of school with confidence in her habits and a plan for the first, second, third, and Nth day of the first year of teaching. In addition to mastering the content they need time mastering the equipment – using, fixing, etc.

2. *Coursework on the learning and teaching of physics*

A quick read through subsection III.D.2 illustrates that the specialized knowledge required to teach physics is not the content focus of even the best reformed physics courses. Personal experience in content courses that model research-based instructional practices may be very valuable in improving students' conceptual understanding and problem-solving, and in certain cases, may even improve students' physics expectations [135]. But having had good teaching modeled does not guarantee that the PSTs in a physics course will be able to then lead others to learn effectively. Specialized courses in physics learning and teaching are required [43]. As a matter of fact, a sequence of such courses is required because habits are unlikely to develop as a result of participation in a single course. (For examples of such courses, see [12] and [139–142].)

The emphasis of these specialized courses is to help PSTs understand and reflect on the ecology of physics ideas and the evidentiary and reasoning chains that have led us to believe what we believe (specifically, experimental testing of explanations, models, hypotheses), learn the fine structure of student ideas that they can anticipate in the topics they will be teaching, learn topic-specific questions or tasks that research has shown to be promising in opening up student discourse, learn the affordances and limitations of different technical representations so that they can recognize physics productivity in students' own spontaneous representations, learn how to assess what is of value to the physics community, etc. [84] Recent research in teacher knowledge shows that it is not subject- but topic-specific, thus the coursework should provide the PSTs with the opportunity to examine every (or as many as possible) physics topics from the teaching point of view, not just have an example from one topic. In this sense, the courses envisioned here work hand-in-hand with the clinical experiences, with a temporal sequencing of courses/practica that front loads what is absolutely needed and no more, and then introduces additional pieces on a need-to-know basis, over a period of time that is sufficiently long for PSTs to integrate knowledge with practice through the exercise of productive habits.

An important element of such courses should be microteaching. Microteaching is a technique used in teacher preparation, whereby PSTs teach a K-12 lesson to their peers, who act as students. Such practice can be extremely useful. PSTs learn to plan a lesson, to interact with students, and to reflect on the lesson after its implementation [45]. This structure also allows multiple opportunities for the mentor (i.e., the course instructor) to provide formative feedback to the PST, and to give the PST an opportunity to revise a particular moment of the lesson and "replay" it. Microteaching is similar to using a flight simulator in pilot training. For pre-service physics teachers microteaching is especially useful (in addition to the general benefits it provides teachers of all subjects) as they have an opportunity to interact with equipment, see the constraints that available equipment puts on their plans and borrow/repair it if needed.

### 3. *Physics teaching community of practice*

A rich and diverse community of practice comprised of the master teacher(s), program faculty and staff--including university supervisors, the PSTs, undergraduate and graduate students interested in teaching and learning, and in-service teachers (especially program graduates) serves multiple goals of the program [45,59,138,143,144]. The community continues to serve as a complex of positive feedback loops as it reinforces habits and common values, as a context for feedback from the field that informs changes to the program, as a social network that increases chances that a teacher candidate will be placed in an instructional environment that embodies the program's vision, and, most importantly for the in-service teachers in the group, as a safe environment in which they can share dilemmas of practice and stay culturally connected to the physics enterprise that inspired them in the first place. The solitary nature of physics teaching and the availability of other career choices to holders of a physics degree are peculiarities of physics teaching that urgently call for special attention to this aspect of a physics teacher education program.

To the best of our knowledge, there exist only a handful of programs in the U.S. that have engendered this kind of community. Two such programs are described in [145]; one is [45], and another is [108]. A third such program is described in [146].

It is in the interwoven nature of the three components (apprenticeship-based clinical practice, coursework on the learning and teaching of physics, and physics teaching community of practice) that the knowledge, skills, dispositions, and the related habits are developed. In addition, a strong sense of identity and enthusiasm about teaching physics is cemented and culturally propagated.

### B. "As a physics teacher educator, what should I do tomorrow?"

The above recommendations might seem overwhelming. A physics teacher educator might ask: What can I do tomorrow so that my program assists my students in the development of the habits described in this paper? In other words, how can we use the above recommendations to help improve currently existing programs? We suggest the following approach.

The starting point is self-assessment of your current program to determine the degree to which the program develops desired habits. The next step is assessment of recent graduates to get to know what is happening in their classrooms and to compare the patterns that emerge to the self-assessment that you conducted. Find the overlapping areas that need improvement and focus first on those that are the easiest to tackle. If we are to use physics language metaphorically to describe this process, we might consider identifying the forces that are exerted on your PSTs and establish which of those forces are easiest to change. Suppose, for example, that you find that the coursework does not help your PSTs develop the necessary physics or physics for teaching knowledge but the number of physics PSTs in the program does not allow you to add teaching methods courses just for the physics PSTs. You might think of adding independent study courses and utilize recently created Physics Teaching Modules [147].

## VI. SUMMARY AND RESEARCH AGENDA FOR THE FUTURE

The goal of this paper has been to provide a conceptual model of physics teacher preparation that explains existing features of successful programs. The DHAC model as described above accomplishes this but in addition to explaining empirically found features, it is also consistent with theoretical frameworks by Ball and Cohen [26] and Hammerness et al. [32] that we discussed in Section II.

The foundation of Ball and Cohen's [26] framework is continuous inquiry into teaching process through practice and reflection. Hammerness et al. [32] consider community to be the foundation of teacher preparation. The community helps PSTs develop "a vision for their practice"; the required knowledge and dispositions; practical skills for the application of knowledge and dispositions; and finally tools "that support their efforts" (p. 385).

The DHAC model, which considers the habits of mind, practice and maintenance the ultimate goal of teacher preparation, accounts for the above-summarized mechanisms by considering both continuity and immersion in a community necessary for habit formation and maintenance.

The next step in our quest towards a theory of teacher preparation would be to make predictions about the graduates of the programs that focus on habit development and conduct studies investigating experimentally these specific predictions. In addition, a host of further research avenues open up and point in turn toward the need for new theory-building. For instance, although the whole model is guided by PER results, there is a need to determine the effect of the program on precollege student learning (by which we mean the whole gamut of physics learning, not just conceptual understanding and problem solving) and on precollege student physics identity formation. To get a handle on these questions we need to operationalize how to document fidelity of implementation of the model. Before fidelity of implementation is studied, however, we need to develop research-validated instruments to measure the specialized, physics-specific knowledge, skills, and dispositions that we have outlined. To be successful in this endeavor, we need in turn tighter theoretical constructs for these terms and ways to apply these constructs to the learning of physics. We provided some examples of the above constructs in the manuscript, but we feel that observational studies documenting the enactment of these constructs specifically in a physics classroom will be extremely helpful for those who prepare physics teachers. We need to create online libraries of annotated videos of high quality physics lessons in which instantiation of knowledge, skills, and dispositions are noted so that teacher educators and PSTs can learn from those videos. Furthermore, we need research programs that seek to document and understand the various ways in which the physics teaching communities of practice support the development, maintenance, and honing of the requisite habits.

It is our sincere hope that the PER community will pursue some of these possible research agenda. There is a national and international need to bring to bear on physics teacher education the PER perspectives, tools, values, and habits that have contributed so much to reformed physics instruction.

Acknowledgements: We are indebted to David Meltzer for his numerous suggestions for the improvement of the paper.## VII. REFERENCES

[1] David E. Meltzer, Monica Plisch, and Stamatis Vokos, editors, *Transforming the Preparation of Physics Teachers: A Call to Action. A Report by the Task Force on Teacher Education in Physics (T-TEP)* (2012).

[2] D. E. Meltzer and V. K. Otero, A brief history of physics education in the United States, Am. J. Phys. **83**, 447 (2015).

[3] National Board for Professional Teaching Standards, *Science Standards,* http://boardcertifiedteachers.org/sites/default/files/EAYA-SCIENCE.pdf (2014).

[4] Council for the Accreditation of Educator Preparation, *CAEP Accreditation Standards,* http://caepnet.org/~/media/Files/caep/standards/final-Board-Amended-20150612.pdf?la=en (2015).

[5] National Task Force On Teacher Education In Physics, *Preparing High School Physics Teachers to Build a 21st Century STEM-Capable Workforce.* https://www.aapt.org/Resources/policy/upload/Statement-on-Teacher-Preparation-Policy-Layout-All-Comments-Accepted.pdf (2012).

[6] P. Callahan, B. Cannon, E. Chesick, J. Mackin, S. Mandel, and C. Wenning, *The Role, Education, Qualifications, and Professional Development of Secondary School Physics Teachers.*
24